\documentclass[a4paper,twocolumn,superscriptaddress,11pt,accepted=2018-11-27]{quantumarticle}
\pdfoutput=1
\usepackage[utf8]{inputenc}
\usepackage[english]{babel}
\usepackage[T1]{fontenc}
\usepackage{amsmath}
\usepackage{hyperref}
\usepackage{soul}

\usepackage{tikz}
\usepackage{lipsum}

\begin{document}

\title{Experimental investigation of high-dimensional quantum key distribution protocols with twisted photons}
\date{\today}
\author{Fr\'ed\'eric Bouchard}
\email{fbouc052@uottawa.ca}
\affiliation{Department of physics, University of Ottawa, Advanced Research Complex, 25 Templeton, Ottawa ON Canada, K1N 6N5}
\author{Khabat Heshami}
\affiliation{National Research Council of Canada, 100 Sussex Drive, Ottawa ON Canada, K1A 0R6}
\author{Duncan England}
\affiliation{National Research Council of Canada, 100 Sussex Drive, Ottawa ON Canada, K1A 0R6}
\author{Robert Fickler}
\affiliation{Department of physics, University of Ottawa, Advanced Research Complex, 25 Templeton, Ottawa ON Canada, K1N 6N5}
\author{Robert~W.~Boyd}
\affiliation{Department of physics, University of Ottawa, Advanced Research Complex, 25 Templeton, Ottawa ON Canada, K1N 6N5}
\affiliation{Max-Planck-Institut f\"{u}r die Physik des Lichts, Staudtstra\ss e 2, 91058~Erlangen, Germany}
\affiliation{Institute of Optics, University of Rochester, Rochester, NY 14627, USA}
\author{Berthold-Georg Englert}
\affiliation{Centre for Quantum Technologies, National University of Singapore, 3 Science Drive 2, Singapore 117543, Singapore}
\affiliation{Department of Physics, National University of Singapore, 2 Science Drive 3, Singapore 117542, Singapore.}
\affiliation{MajuLab, CNRS-UNS-NUS-NTU International Joint Unit, UMI 3654, Singapore.}
\author{Luis L. S\'anchez-Soto}
\affiliation{Max-Planck-Institut f\"{u}r die Physik des Lichts, Staudtstra\ss e 2, 91058~Erlangen, Germany}
\affiliation{Departamento de \'{O}ptica, Facultad de F\'{\i}sica, Universidad Complutense, 
28040~Madrid, Spain}
\author{Ebrahim Karimi}
\email{ekarimi@uottawa.ca}
\affiliation{Department of physics, University of Ottawa, Advanced Research Complex, 25 Templeton, Ottawa ON Canada, K1N 6N5}
\affiliation{Max-Planck-Institut f\"{u}r die Physik des Lichts, Staudtstra\ss e 2, 91058~Erlangen, Germany}
\affiliation{Department of Physics, Institute for Advanced Studies in Basic Sciences, 45137-66731 Zanjan, Iran.}

\begin{abstract}
 \begin{abstract}
Quantum key distribution is on the verge of real world applications, where perfectly secure information can be distributed among multiple parties. Several quantum cryptographic protocols have been theoretically proposed and independently realized in different experimental conditions. Here, we develop an experimental platform based on high-dimensional orbital angular momentum states of single photons that enables implementation of multiple quantum key distribution protocols with a single experimental apparatus. Our versatile approach allows us to experimentally survey different classes of quantum key distribution techniques, such as the 1984 Bennett \& Brassard (BB84), tomographic protocols including the six-state and the Singapore protocol, and {to investigate, for the first time, a recently introduced differential phase shift (Chau15) protocol using twisted photons}. This enables us to experimentally compare the performance of these techniques and discuss their benefits and deficiencies in terms of noise tolerance in different dimensions. 
\end{abstract}
\end{abstract}

\section{Introduction}
Quantum cryptography allows for the broadcasting of information between multiple parties in a perfectly secure manner under the sole assumption that the laws of quantum physics are valid~\cite{gisin:02}.  Quantum Key Distribution (QKD)~\cite{bennett:84,scarani:09} is arguably the most well-known and studied quantum cryptographic protocol to date. Other examples are quantum money~\cite{wiesner:83} and quantum secret sharing~\cite{hillery:99}. In QKD schemes, two parties, conventionally referred to as \emph{Alice} and \emph{Bob}, exchange carriers of quantum information, typically photons, in an untrusted quantum channel. An adversary, known as \emph{Eve}, is granted full access to the quantum channel in order to eavesdrop on Alice and Bob's shared information. It is also assumed that Eve is only limited by the laws of physics and has access to all potential future technologies to her advantage, including optimal cloning machines~\cite{scarani:05}, quantum memories~\cite{simon:10}, quantum non-demolition measurement apparatus~\cite{werner:93}, and full control over the shared photons. In particular, the presence of Eve is revealed to Alice and Bob in the form of noises in the channel. It is the goal of QKD to design protocols for which secure information may be transmitted even in the presence of noises~\cite{bennett:95}. For quantum channels with high levels of noises, it has been recognized that high-dimensional states of photons constitute a promising avenue for QKD schemes, due to their potential increase in noise tolerance with larger encrypting alphabet~\cite{bechmann:00b,cerf:02}. However, this improvement comes at the cost of generating and detecting complex high-dimensional superpositions of states, which may be a difficult task.

Orbital angular momentum (OAM) states are associated with helical phase fronts for which a quantized angular momentum value of $\ell \hbar$ along the photons propagation direction can be ascribed, where $\ell$ is an integer and $\hbar$ is the reduced Planck constant~\cite{allen:92}. Any arbitrary superposition of OAM states can be straightforwardly realized by imprinting the appropriate transverse phase and intensity profile on an optical beam, which is typically done by displaying a hologram onto a spatial light modulator (SLM)~\cite{heckenberg:92,bolduc:13,forbes:16}. OAM-carrying photons, also known as twisted photons, have  been recognized to constitute useful carriers of high-dimensional quantum states for quantum cryptography~\cite{groblacher:06,mafu:13,mirhosseini:15}, quantum communication~\cite{dambrosio:12,vallone:14,krenn:15,sit:17} and quantum information processing~\cite{cardano:15,cardano:16,cardano:17,babazadeh:17,erhard:17}. The flexibility in preparation and measurement of twisted-photon states enables us to create and use a single experimental setup for implementing several QKD protocols that offer different advantages, such as efficiency in secure bit rate per photon or noise tolerance for operation over noisy quantum channels. Here, we use OAM states of photons to perform and compare high-dimensional QKD protocols such as the 2-, 4- and 8-dimensional BB84~\cite{bennett:84}, tomographic protocols~\cite{bruss:98,liang:03,englert:04} using mutually unbiased bases (MUB)~\cite{durt:10} and Symmetric Informationally Complete (SIC) Positive Operator-Valued Measures (POVMs)~\cite{renes:04}, and, {for the first time}, the 4- and 8-dimensional \emph{Chau15} protocols {using twisted photons}. We finally demonstrate applications in full characterization of the quantum channel through quantum process tomography~\cite{chuang:97}.

\section{Theoretical background}
Let us first start by briefly reviewing the BB84 protocol, which was introduced in 1984 by Bennett and Brassard~\cite{bennett:84}. In this protocol, Alice uses qubits to share a bit of information (0 or 1) with Bob, while using two different MUB~\cite{durt:10}. This QKD protocol relies on the \emph{uncertainty principle}, since a measurement by Eve in the wrong basis will not yield any useful information for herself. However, this also means that half of the time, Alice and Bob will perform their generation/detection in the wrong basis. This is known as sifting: Alice and Bob will publicly declare their choices of bases for every photon sent and only when their bases match will they keep their shared key. On average, Alice and Bob will only use half of their bits in their shared \emph{sifted key}. {Nevertheless, in the infinite key limit, the sifting efficiency of several protocols, such as the BB84 and the six-state protocol, can approach 1, by making the basis choice extremely asymmetric~\cite{lo:05}.} In addition to sifting, Alice and Bob's shared key will be further reduced in size at the final stage of the protocol when performing error correction (EC) and privacy amplification (PA)~\cite{bennett:95}. In the case of BB84 in dimension 2, the number of bits of secret key established per sifted photon, defined here as the secret key rate $R$, is given by the following expression,

\begin{equation}
R=1-2 h(e_b),
\label{eq:bb84}
\end{equation}
where $e_b$ is the quantum bit error rate (QBER) and $h(x):=-x \log_2 (x) - (1-x) \log_2 (1-x)$ is the Shannon entropy. From this equation, we find that the secret key rate becomes negative for $e_b>0.11$. Hence, if Alice and Bob only have access to a quantum channel with a QBER larger than 0.11 to perform the BB84 protocol, they will not be able to establish a secure key regardless of sifting and losses. Due to this limitation, it has been the goal of many research efforts to come up with QKD protocols that are more error tolerant. One of the first proposed QKD protocols that was aimed at extending the $0.11$-QBER threshold, is the so-called \emph{six-state} protocol~\cite{bruss:98}. The six-state protocol is an extension of the BB84 in dimension 2, where all three MUB are used. Indeed, it is known that in dimension 2, there exists precisely 3 MUB. In the general case of a $d$-dimensional state space, the number of MUB is $(d+1)$, given that $d$ is a power of a prime number~\cite{durt:10}. Moreover, it is known that there exist at least three MUB for any dimension~\cite{durt:10} such that these 3 MUB can be used to increase error thresholds in dimensions which are not a power of a prime number~\cite{bradler:16,ding:17}. Of course, this has the drawback of decreasing the efficiency of obtaining sifted data from $1/2$ to $1/3$. Nevertheless, by simply adding another basis to the encoding measurement scheme, the QBER threshold now increases to $0.126$, as can be deduced from the secret key rate of the six-state protocol~\cite{bruss:98} 

\begin{equation}
R = 1 - h \left( \frac{3}{2} e_b \right) - \frac{3}{2} e_b \log_2 (3) .
\end{equation}
\begin{figure*}[t]
	\begin{center}
	\includegraphics[width=2\columnwidth]{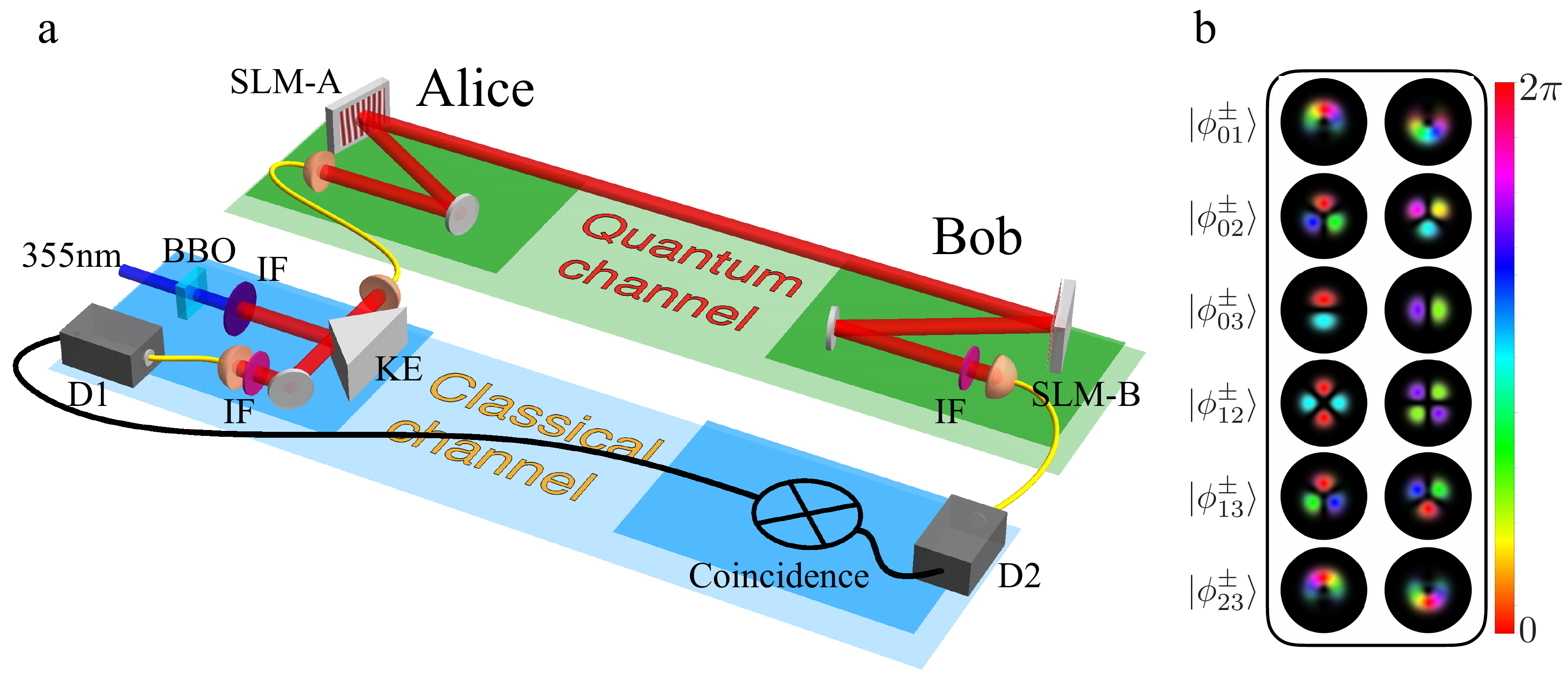}
	\caption[]{(a) Simplified experimental setup. Alice generates pairs of single photons using spontaneous parametric downconversion in a type-I $\beta$-barium borate (BBO) crystal. The pump wavelength (355~nm) is filtered using an interference filter (IF) and the photons are separated using a knife edge (KE) in the far-field of the crystal. The spatial modes of the photon pairs are filtered using single mode optical fibres making the photons completely separable. Alice imprints a state onto her signal photon using a holographic technique by means of a spatial light modulator (SLM-A). The photon is then sent to Bob through the quantum channel. Bob measures the photon's state using a phase flattening technique with his SLM-B followed by a single mode optical fibre. Moreover, Alice locally measures the idler photon and sends timing information to Bob via an electric signal over the classical channel. Our experimental configuration allows us to test different protocols by changing the holograms displayed on the SLMs using the same experimental apparatus without intermediate adjustments.  Thus, we are able to compare the different strategies in a systematic manner. (b) States employed in the Chau15 $(N=4)$ protocol. The phase (Hue colour) is shown modulated by the intensity profile of the beam.}.
	\label{fig:exp}
	\end{center}
\end{figure*}

Another avenue to increase error tolerability in QKD is to use high-dimensional quantum states, also known as \emph{qudits}~\cite{genovese:08}. This may be intuitively understood from the fact that the presence of an optimal cloning attack leads to larger signal disturbance in higher-dimensional QKD schemes~\cite{cerf:02,bouchard:17}. The BB84 protocol may be extended here by using qudits. The adoption of high-dimensional quantum systems has two distinct benefits: (i) an increase of the error-free key rate per sifted photons to  a value of $R=\log_2(d)$; (ii) an increase in the maximum tolerable QBER, i.e. the error threshold for $R=0$. For the simple case of a $d$-dimensional BB84 protocol, the secret key rate is given by~\cite{sheridan:10}, 
\begin{equation}
R=\log_2(d)-2 h^{(d)}(e_b),
\end{equation}
where $h^{(d)}(x):=-x \log_2 (x/(d-1)) - (1-x) \log_2 (1-x)$ is the $d$-dimensional Shannon entropy. Furthermore, it is also possible to extend the \emph{six-state} protocol to higher dimensions by employing all $(d+1)$ MUB, assuming that $d$ is a power of a prime number, where the secret key rate is given by,
\begin{equation}
R=\log_2(d)- h^{(d)}\left(\frac{d+1}{d} e_b\right)-\frac{d+1}{d} e_b \log_2 (d+1).
\end{equation}

This type of QKD scheme is also known as tomographic QKD~\cite{liang:03,englert:04}, where all measurements, including the sifted ones, are used to perform quantum state tomography (QST). In particular, MUB are closely related to the problem of quantum state tomography, where projections over all the states of every MUB, although redundant, yields a full reconstruction of the state's density matrix~\cite{dambrosio:13}. Following similar ideas, the \emph{Singapore} protocol has been proposed using SIC-POVMs~\cite{englert:04}, as they are known to be the most efficient measurements to perform QST. The Singapore protocol may be equivalently performed in a prepare-and-measure or an entanglement-based scheme, similar to the analogy between BB84 and Ekert~\cite{ekert:91}, respectively. Moreover, this QKD protocol may also be extended to higher dimensions~\cite{bent:15} with the major advantage that the SIC-POVMs are believed to exist for all dimensions, including those that are not powers of prime numbers, contrary to MUB.

Another class of QKD protocols using qudits has recently been introduced, in which qudits are used to encode a single bit of information. Although, such protocols primarily benefits from one of the advantages mentioned earlier, i.e. an increase in the QBER threshold, they have proven to be interesting and advantageous due to a simplified generation and measurement of the states. Their main drawback is the fact that at a null QBER, the key rate per sifted photons is never larger than $R=1$. An example of such a protocol is the \emph{Differential Phase Shift} (DPS) QKD protocol~\cite{inoue:02}. The information is encoded by Alice in the relative phase of a superposition of all states then sent over the quantum channel. Bob may then measure the relative phase by detection of the different phases using an interferometric apparatus. In particular, the advantage of the 3-dimensional DPS scheme is in the higher sifting efficiency in comparison to BB84 {in the finite key limit}. An extension of the DPS protocol is the \emph{Round-Robin Differential Phase Shift} (RRDPS) protocol~\cite{sasaki:14} where Bob's interferometric apparatus is slightly modified. This modification results in a bound on Eve's leaked information removing the need to monitor signal disturbance (QBER) for performing privacy amplification. Nevertheless, the qudits employed in the RRDPS QKD protocol consist of a superposition of $d$ states, which may pose some practical limitations in experimental implementations as $d$ becomes larger. However, the RRDPS scheme has recently been demonstrated experimentally using twisted photons~\cite{Bouchard:18b}. The recently introduced \emph{Chau15} protocol~\cite{chau:15,chau:17} addresses this problem as it uses ``qubit-like'' superpositions where only two states of the $d$-dimensional space are employed. More specifically, the information is encoded in the relative phase of a qubit-like state of the form $|\phi_{ij}^{\pm}\rangle=(|i\rangle \pm |j\rangle)/\sqrt{2}$ with states in a $2^n$-dimensional Hilbert space with $n\geq 2$. This protocol will be explained in more details in the discussion section.

\section{Experimental setup}
We implement a prepare-and-measure QKD scheme at the single-photon level using the OAM degree of freedom of photons, see Fig.~\ref{fig:exp} (a). The single photon pairs, namely \emph{signal} and \emph{idler}, are generated by spontaneous parametric downconversion (SPDC) at a type I $\beta$-barium borate (BBO) crystal. The nonlinear crystal is pumped by a quasi-continuous wave ultraviolet laser operating at a wavelength of 355~nm. The generated photon pairs are coupled to single-mode optical fibres (SMF) in order to filter their spatial modes to the fundamental mode; i.e., Gaussian. Following the SMF, a coincidence rate of 30~kHz is measured within a coincidence time window of 2~ns. The heralded signal photon is sent onto SLM-A corresponding to Alice's generation stage. {The SLM (X10468-07, Hamamatsu) are electronically controlled nematic liquid crystal devices with $792 \times 600$ pixels, a refresh rate of 60~Hz and a diffraction efficiency in excess of 70~\%}. The OAM states are produced using a phase-only holography technique~\cite{bolduc:13}. Due to the versatility of SLMs, any OAM superposition states of single photons may be produced, hence covering a large possibility of QKD schemes. Alice's heralded photon is subsequently sent over the untrusted quantum channel. Upon reception of the photon, Bob uses his SLM-B followed by a SMF to perform a projection over the appropriate states for a given protocol. In order to do so, Bob uses the phase-flattening technique to measure OAM states of light~\cite{mair:01,qassim:14}. If the incoming photon carried the OAM mode corresponding to Bob's projection, the phase of the mode is flattened and the photon will couple to the SMF. This verification-type measurements can further reduce the sifting rate, unless the protocols requires only a binary measurement and ``no-click'' events are included. Coincidences are recorded using single photon detectors (ID120-500-800nm, ID Quantique) with a dark count rate of less than 50~Hz and a measured deadtime of approximately 400~ns. For an integration time of 10~s and a coincidence time window of 2~ns,  approximately 14,000 coincidences are recorded when both SLM-A and SLM-B are set to the same mode. This corresponds to a transmission of approximately 5~\%. Finally, in order to evaluate the single photon nature of the SPDC source, we build a Hanbury-Brown and Twiss interferometer in order to measure the degree of second-order coherence $g^{(2)}(0)$ for our single photon source in a three-detector measurement configuration. The signal photon is sent to a beam splitter and subsequently measured at both output ports of the beam splitter, i.e. detectors $A$ and $B$, while the idler photon is measured directly at detector $C$. The experimentally determined second-order coherence is given by $g^{(2)}(0) = (N_{ABC} N_C) / (N_{AC } N_{BC})$, where $N_{ABC}$ is the three-fold coincidence rate among detectors $A$, $B$ and $C$, $N_C$ is the single count rate at detector $C$, $N_{AC}$ is the coincidence rate between detector $A$ and $C$ and $N_{BC}$ is the coincidence rate between $B$ and $C$. We obtained an experimental value of $g^{(2)}(0) = 0.015 \pm 0.004$, where the experimental uncertainty is calculated assuming Poissonian statistics. {For the case of BB84, the knowledge of the $g^{(2)}(0)$ and the efficiency of the source is sufficient to characterize the effect of multiphoton events on the secret key rate~\cite{waks:02}. Following the analysis of~\cite{schiavon:16}, the secret key rate given in Eq.~\ref{eq:bb84} becomes:

\begin{equation}
R = \left( 1 -\Delta \right) \left( 1- h\left( \frac{e_b}{1-\Delta} \right) \right) - h\left(e_b \right),
\end{equation}

where $\Delta = P_m / Q$ is the multiphoton rate, $P_m = 1-P_0 - P_1$ is the probability of having more than one photon in a pulse, $Q=\sum_{n=0}^\infty Y_n P_n$ is the gain, $Y_n$ is the yield of an $n$-photon signal and $P_n$ is the probability of having $n$ photons in a pulse. The probability of creating a multiphoton state, $P_m$, is upper bounded by $g^{(2)}(0)$, i.e. $P_m \leq \mu^2 g^{(2)}(0) / 2$, where $\mu$ is the mean photon number in a pulse. From the experimental parameters mentioned above, we obtain a gain, a mean photon number, and a multiphoton rate of $Q=10^{-5}$, $\mu = 3 \times 10^{-4}$, and $\Delta = 4 \times 10^{-5}$, respectively, which will have a negligible effect on the secret key rate. For instance, in the case of BB84 in dimension 2, the secret key rate considering multiphoton events is reduced by an amount on the order of $10^{-5}$ bits per sifted photon.}

\begin{figure}[t]
	\begin{center}
	\includegraphics[width=1.0\columnwidth]{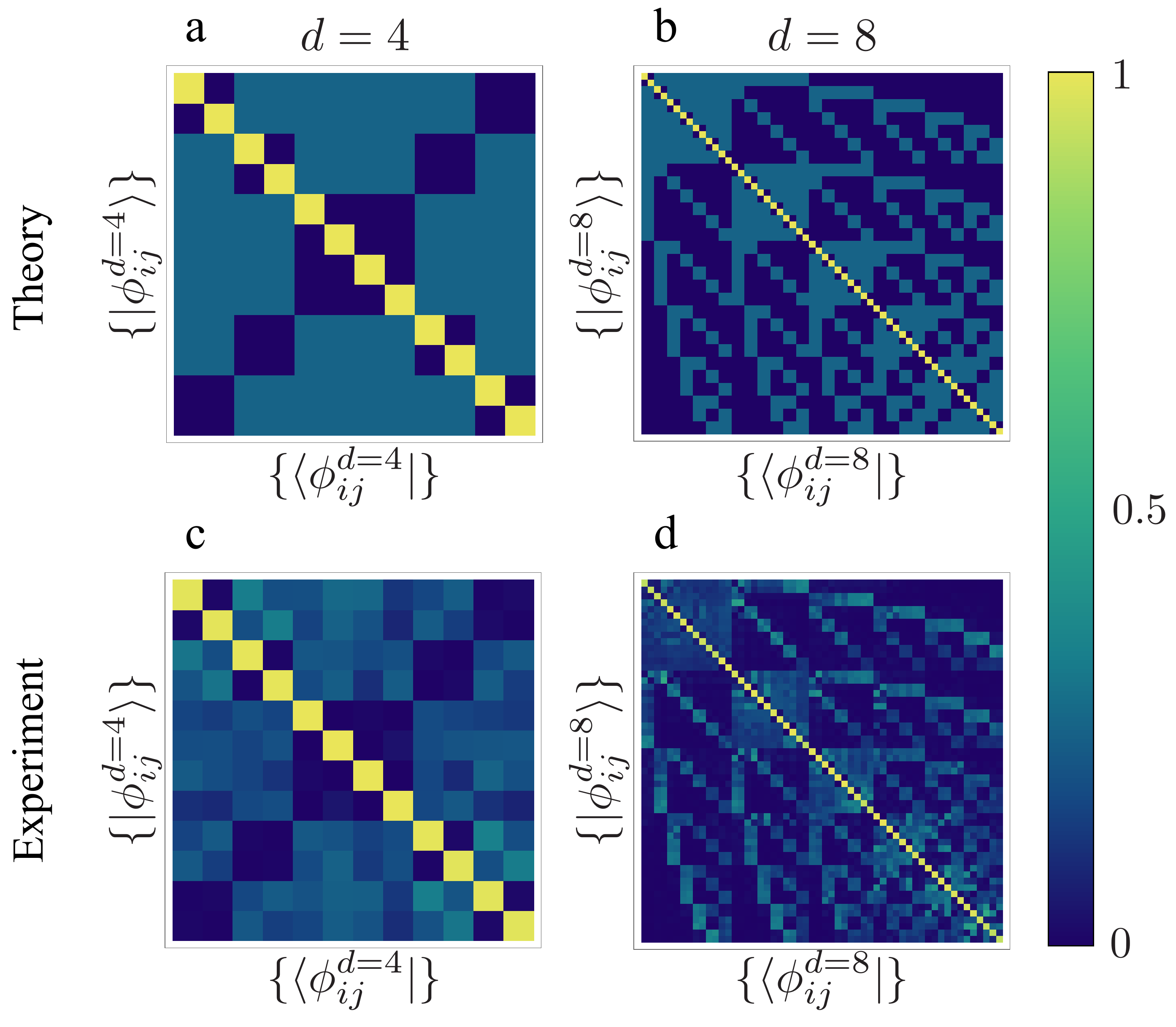}
	\caption[]{Results for the Chau15 protocol. (a)-(b) Theoretical probability-of-detection matrices for the Chau15 protocol in dimension $d=4$ and $d=8$, respectively. Rows are given by states $| \phi_{ij} ^\pm \rangle \in \{ |\phi_{1,2}^+ \rangle, |\phi_{1,2}^- \rangle, |\phi_{1,3}^+ \rangle, ... , |\phi_{3,4}^- \rangle \}$ sent by Alice, whereas columns corresponds to sate projections, $\langle \phi_{ij}^\pm |$, by Bob. (c)-(d) Experimentally measured probability-of-detection matrices for the Chau15 protocol in dimension $d=4$ and $d=8$. The sifted data corresponds to the on-diagonal $2\times2$ blocks. The remaining of the probability-of-detection matrix may be used to evaluate the dit error rate. }
	\label{fig:chau15}
	\end{center}
\end{figure}

\section{Results and Discussion}
\subsection{Chau15 protocol} 
At first, we perform the recently introduced Chau15 protocol, which is a qudit-based prepare-and-measure QKD scheme, where information is encoded in a qubit-like state of the form $|\phi_{ij}^{\pm}\rangle=(|i\rangle \pm |j\rangle)/\sqrt{2}$ with states in a $2^n$-dimensional Hilbert space with $n\geq 2$; see~\cite{chau:15,chau:17,wang:17}.
The protocol starts with Alice randomly selecting $i$, $j$, and $s$, where $\{i,~j \} \in  \mathrm{GF}(d=2^n)$, $\mathrm{GF}(d)$ being the Galois field, and $s=\pm 1$. Then, Alice prepares and sends the state $\left(|i\rangle +(-1)^s |j\rangle\right)/\sqrt{2}$ over an untrusted channel to Bob. Upon reception, Bob randomly selects $i',~j'\neq i' \in \mathrm{GF}(d)$ and measures the state along $(|i'\rangle \pm |j'\rangle)/\sqrt{2}$. By announcing $(i,j)$ and $(i',j')$ through a classical channel, Alice and Bob can establish a raw bit sequence (key) from those events where $(i,j)=(i',j')$ and by keeping a record of $s$. 

As discussed in~\cite{chau:17}, the performance of the scheme can be assessed through two sets of parameters. The first is the bit error rate of the sifted raw key ($e_{\text{raw}}$). The second is the average bit error rate and the average dit error rate associated with mismatch between preparation and measurement basis states. The average bit and dit error rates are estimated by averaging probabilities of the qudit states undergoing operations of $X_{u}Z_{v}$ in the insecure quantum channel, where $X_u|i\rangle=|i+u\rangle$, $Z_v|i\rangle=(-1)^{\text{Tr}(vi)}|i\rangle$, and $\text{Tr}(i)=i+i^{2}+i^{4}+...+i^{d/2}$; see~\cite{chau:17} for more details. These parameters can be extracted from the experimental joint probability measurements. Alice and Bob respectively prepare and measure $|\phi_{ij}^{\pm}\rangle=(|i\rangle \pm |j\rangle)/\sqrt{2}$, where $|i\rangle$ and $|j\rangle$ $(i\neq j)$ are pure OAM states in a Hilbert space of dimension $d=4,~8$, see Fig.~\ref{fig:exp} (b). Using the OAM, Alice and Bob choose $i,j \in \{\ell = -2,-1,1,2\}$ for $d=4$ and $i,j \in \{\ell = -4,-3,-2,-1,1,2,3,4 \}$ for $d=8$, where the $\ell=0$ state has been omitted to make the states symmetric. In Fig.~\ref{fig:chau15}, we show theoretical and experimental probability-of-detection matrices obtained from a Chau15 QKD protocol for the cases of $d=4$ and $d=8$. After sifting, Alice and Bob are left with the on-diagonal $2\times2$ blocks of the presented probability-of-detection matrices. 

In the case of $d=4$, we obtained an average bit error rate of $e_b^{(d=4)} = 0.778~\%$, and average dit error rate $e_d^{(d=4)} = 3.79~\%$. This results in an asymptotic secure key rate of $R^{(d=4)} = 0.8170$ bit per sifted photon. For the case of $d=8$, we obtained experimental values for the average bit error rate, average dit error rate and asymptotic secure key rate of $e_b^{(d=8)} = 3.11~\%$, $e_d^{(d=8)} = 0.82~\%$ and $R^{(d=8)} = 0.8172$, respectively. Note that the probability of obtaining sifted data in the Chau15 protocol is given by $2/(d^2-d)$ compared to the fixed sifting rate of 1/2 for the BB84 protocols in all dimensions. As we will see in the following sections, the Chau15 protocol does not perform well in the low-error case compared with other QKD protocols. Moreover, the unfavourable scaling of the sifting with dimensionality greatly affects the overall secure key rate (sifting included) {in the finite key limit}. Hence, the advantage of the Chau15 scheme is in the high-error scenario. In particular, given a small enough dit error rate, bit error rates of up to $e_b^\mathrm{max}=50~\%$ may be tolerated. In the case of OAM states of light, this does not represent a clear advantage since bit and dit error rates will, in general, be the result of similar error sources, e.g. misalignment, turbulence or optical aberration. However, for other kinds of high-dimensional states of light, such as time-bins, the distinction between bit and dit error rates is less ambiguous and the Chau15 may then be used to its full potential.

\subsection{BB84 protocol} 
We now go on to compare the new Chau15 scheme to established protocols. The same experimental setup is used at first to perform the BB84 (2 MUB) protocol in dimension $d=2,4$ and 8. In the BB84 protocols, the first basis is given by the logical pure OAM basis, i.e. $|\psi_i \rangle \in \{-d/2,...d/2\}$ and the second basis is given by the Fourier basis where the states are obtained from the discrete Fourier transform, $|\phi_i \rangle=\frac{1}{\sqrt{d}}\sum_{j=0}^{d-1} \omega_d^{ij} |\psi_i\rangle$, with $\omega_d=\exp (i 2\pi / d)$. The explicit form of the other MUB may be found elsewhere~\cite{durt:10}. Although this works only for prime dimensions, it can be easily extended to composite dimensions. For the BB84 protocol, values of the QBER of $e_b^{d=2}=0.628~\%$, $e_b^{d=4}=3.51~\%$ and $e_b^{d=8}=10.9~\%$ were obtained in dimension 2, 4 and 8, respectively corresponding to secure key rates of $R^{d=2}=0.8901$, $R^{d=4}=1.4500$ and $R^{d=8}=1.3942$. In the low-error case, the BB84 scheme performs very well. This is partly due to the fact that the sifting, i.e. $1/2$, is independent of dimensionality {in the finite key limit}. Interestingly, the BB84 protocol performs better in dimension 4 than it does in dimension 8. Hence, although in the error-free case, larger dimensions result in larger secure key rates, this is not necessarily the case in experimental implementations, due to more complex generations and detections of the high-dimensional OAM states. The nature of the quantum channel may also dictate the optimal dimensionality of the protocol~\cite{Bouchard:18}.
\begin{figure}[h]
	\begin{center}
	\includegraphics[width=1.0\columnwidth]{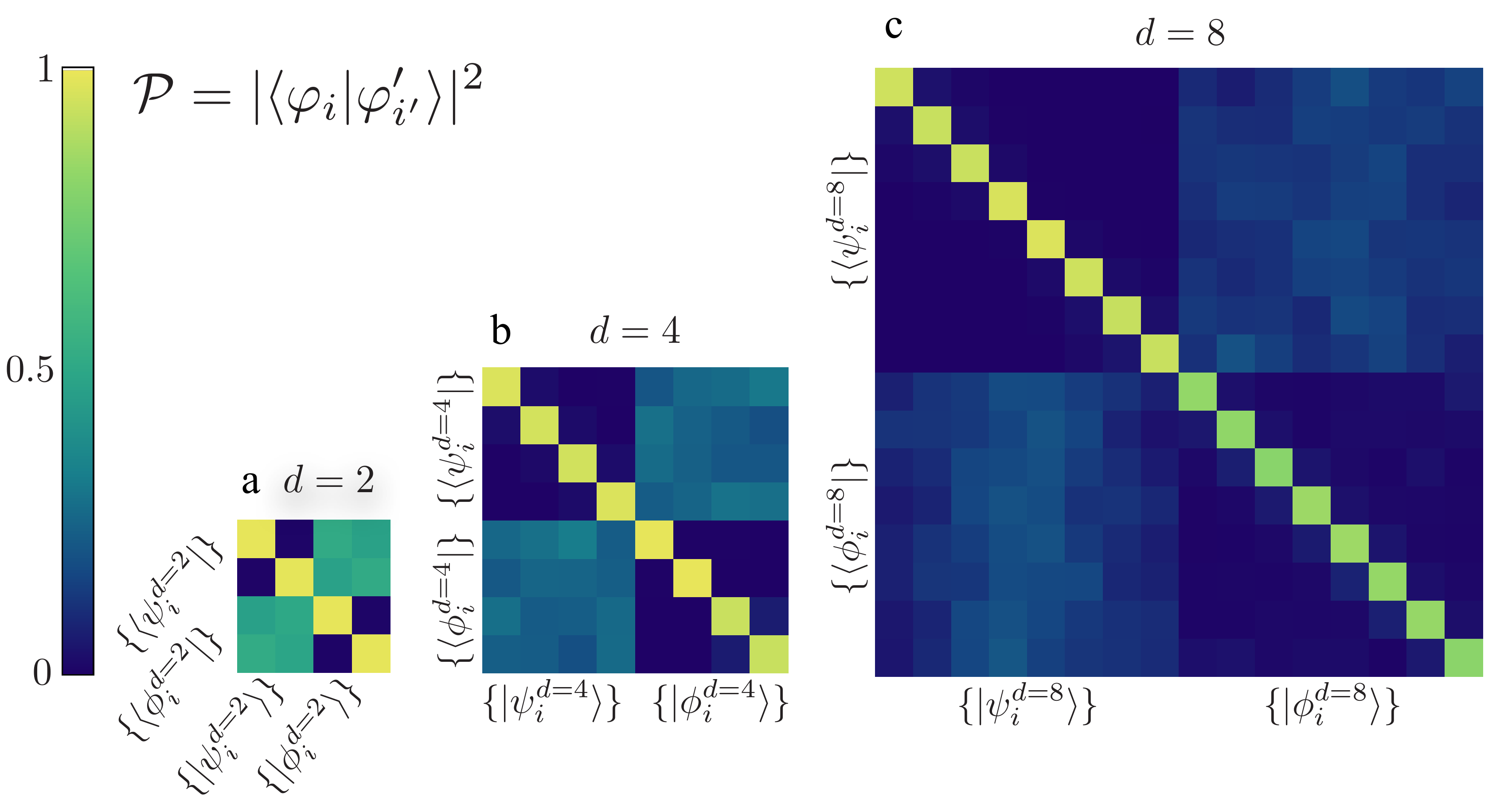}
	\caption[]{BB84 protocol. (a)-(c) Experimentally measured probability-of-detection matrices for the BB84 protocol in dimension $d=2, 4$ and 8, respectively. The rows and columns of the matrices correspond to the states sent and measured by Alice and Bob, respectively. The sifted data corresponds to the on-diagonal $d \times d$ blocks.}
	\label{fig:BB84}
	\end{center}
\end{figure}

\subsection{Tomographic protocols}
\subsubsection{MUB-based protocol}
\begin{figure*}[!htbp]
	\begin{center}
	\includegraphics[width=2\columnwidth]{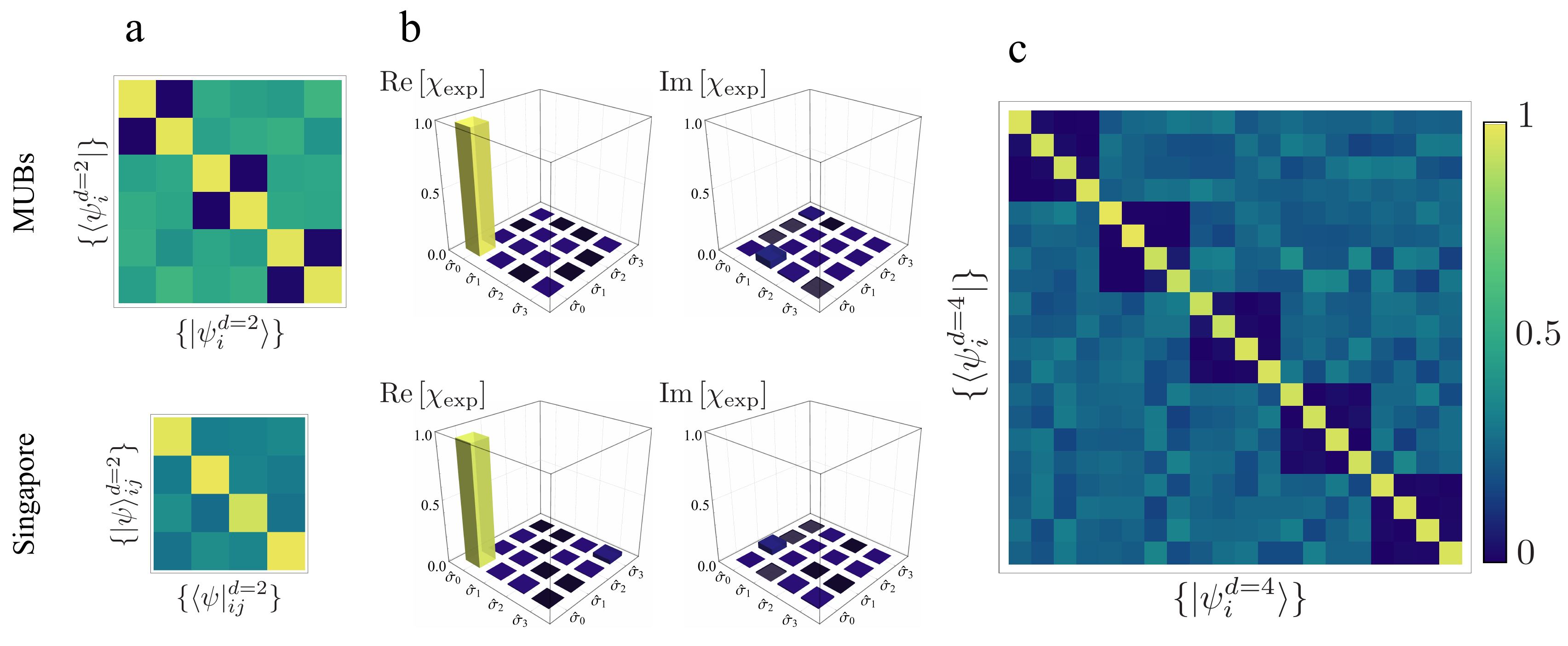}
	\caption[]{Results for the $(d+1)$-MUB and Singapore protocols. (a) Experimentally measured probability-of-detection matrices for the $(d+1)$-MUB and Singapore protocols in dimension $d=2$. The rows and columns of the matrices correspond to the states sent and measured by Alice and Bob, respectively. For the case $(d+1)$-MUB protocols, the sifted data corresponds to the on-diagonal $d \times d$ blocks. (b) Reconstructed process matrix for the six-state (upper) and the Singapore (lower) protocols. (c) Experimentally measured probability-of-detection matrices for the $(d+1)$-MUB protocol in dimension $d=4$. }
	\label{fig:MUBSing}
	\end{center}
\end{figure*}

The six-state protocol in dimension $d=2$ is an extension of the BB84 protocol where all existing MUB are considered. The protocol can be extended to higher dimensions (powers of prime numbers) where all $(d+1)$ MUB are considered. In dimension 2 and 4, we obtained a QBER of $e_b^{d=2,m=3}=0.923~\%$ and $e_b^{d=4,m=5}=3.87~\%$ for the 3-MUB and the 5-MUB protocols, see Fig.~\ref{fig:MUBSing} (a) and (c), corresponding to key rates of $R^{d=2,m=3}=0.8727$ and $R^{d=4,m=5}=1.5316$ bits per sifted photon. In comparison to the BB84 protocol, the $(d+1)$-MUB protocol has a sifting efficiency of $1/(d+1)$, which scales poorly with dimensions. {Nevertheless, in the infinite key limit, the $(d+1)$-MUB approach could exceed the performance of the BB84 protocol by considering an efficient asymmetric basis choice.} Furthermore, this scheme only applies to dimensions that are powers of prime numbers. However, the $(d+1)$-MUB approach consists of a tomographic protocol and in the case of large errors, it will outperform the BB84 scheme. In practical implementations, one could consider an intermediate scenario where the number of MUB considered is between 2 and $(d+1)$ in order to optimize the secure key rate.

\subsubsection{Singapore protocol} Finally, we perform another tomographic QKD protocol based on SIC-POVMs, known as the Singapore protocol. In particular, we use the Weyl-Heisenberg covariant SIC-POVMs elements. Reference vectors $|f\rangle$ have been conjectured to exist in arbitrary dimensions~\cite{renes:04,scott:10} such that SIC-POVMs can be obtained by considering a displacement operator $\hat{D}_{jk}$ acting on the reference vector $|f\rangle$, where
\begin{equation}
\hat{D}_{jk}=\omega_d^{jk/2} \sum_{m=0}^{d-1} \omega_d^{jm} | k + m \rangle \langle m |, 
\end{equation}
and $\omega_d=e^{2\pi i/d}$. The fiducial vectors $|f\rangle$ have been derived numerically and analytically for different dimensions~\cite{renes:04}. The fiducial vector is found such that \mbox{$|\psi_{jk}\rangle =\{\hat{D}_{jk}|f\rangle ,~j,k=0...d-1\}$} are normalized states satisfying $|\langle\psi_{jk}|\psi_{j'k'}\rangle|^2=1/(d+1)$ for $j\neq j'$ and $k\neq k'$. This set of $d^2$ states are then used by Alice and Bob in the prepare-and-measure Singapore protocol. {In dimension 2, the SIC POVMs are explicitly given by

\begin{eqnarray*}
|\psi_{0,0} \rangle &=& \left( 0.888 |0\rangle + 0.325(1-i)|1\rangle \right), \\
|\psi_{0,1} \rangle &=& \left( 0.325(1-i)|0\rangle + 0.888 |1\rangle \right), \\
|\psi_{1,0} \rangle &=& \left( 0.888 |0\rangle - 0.325(1-i)|1\rangle \right), \\
|\psi_{1,1} \rangle &=& \left( 0.325(1+i)|0\rangle - 0.888 i |1\rangle \right).
\end{eqnarray*}

} Using the same experimental apparatus, we perform the Singapore protocol in dimension $d=2$, where a QBER of $e_b=1.23~\%$ was measured, see Fig.~\ref{fig:MUBSing} (a).

Inspired by the Singapore protocol~\cite{englert:04} an iterative key extraction method can be applied to extract a sifted secret key surpassing the 1/3 limit of the six-state protocol. The asymptotic efficiency of the iterative approach have been shown to reach 0.4 which is slightly smaller than the theoretical maximum of 0.415 under ideal conditions~\cite{englert:04}. Here, we use the experimental joint probability matrix to find the experimental mutual information as an upper bound for the key extraction rate. For this purpose, we parametrize $|\psi_m^{A,B}(x)\rangle$ vectors (as Alice's and Bob's experimental preparation and measurement states). We then numerically minimize a maximum likelihood relation of the form $f(x)=\sum_{m,n=1}^{d^2}{\left| |\langle\psi_m^A(x)|\psi_n^B(x)\rangle|^2-|\langle\psi_m|\psi_n\rangle|^2\right|^2}$ to find deviations (errors) in the experimental SIC states of Alice and Bob. These deviations can also be interpreted as errors in the quantum channel. 

The Singapore protocol relies on an anti-correlation between Alice an Bob. In the entanglement-based version of this protocol, this can be achieved by sharing a singlet entangled state. Assuming a singlet state of $|\Psi^{(-)}\rangle_{AB}=\frac{1}{\sqrt{d}}\Sigma_{m=0}^{m=d-1}(-1)^{d-m}|m\rangle|d-m-1\rangle$, we can extract the joint anti-correlated prepare-and-measure probabilities. In dimension $d=2$, this takes the form of $p_{kl} = \text{Tr} [ \hat{\rho}_{AB} (1+\vec{t}_k \cdot \vec{\sigma}_A) (1+\vec{t}_l \cdot \vec{\sigma}_B )]$, where $\vec{t}_k$s are unit vectors denoting SIC states and ${\hat\sigma}_{i}$s are Pauli matrices. This typically deviates from the ideal case with prepare-and-measure probabilities of $p_{kl} =(1-\delta_{kl})/12$. Keep in mind that the Singapore protocol relies on completely symmetric prepare-and-measure probabilities. This symmetrization can be achieved by twirling the calculated probability matrix leading to 
\begin{equation}
p^{\text{exp}}_{kl}=\frac{4-\epsilon}{48}\left(1-\delta_{kl}\right)+\frac{\epsilon}{16}\delta_{kl},
\end{equation}
where $\epsilon = 0.0137$.
The mutual information between Alice and Bob is given by
\begin{equation}\label{I_AB}
I_{AB} = \sum_{k,l=1}^{d^2} p_{kl}\log_2 \left( \frac{p_{kl}}{p_{k}  p_{l}} \right),
\end{equation}
where $p_{k} = \sum_{l=1}^{d^2}p_{kl}$ and $p_{l} = \sum_{k=1}^{d^2}p_{kl}$.
Our approach results in mutual information of $I_{AB}^{d=2}=0.388$ compared to the theoretical maximum of 0.415; surpassing the maximum attainable rate in the six-state scheme. Moreover, protocols analogous to the Singapore protocol have a poor yield for higher-dimensional systems. The mutual information for qubit, qutrit and ququart pairs are respectively given by 0.415, 0.170 and 0.093 bits.

\subsubsection{Quantum process tomography of the QKD channel}
In this subsection we considered the two QKD protocols that offer full tomography capabilities~\cite{chuang:97,ndagano:17}. Quantum state preparation and measurements in all possible MUB states provide an overcomplete set of results that can be used to perform quantum process tomography on the channel. In the Singapore protocol, the SIC POVMs are optimal set of preparation and measurements for process tomography of the channel. Both of these protocols allow one to go beyond a coarse-grained qubit error rate estimation and fully characterize the quantum channel. We use the experimental results for the six-state and Singapore protocols to characterize the quantum channel.

\begin{table*}
\centering
	\begin{tabular}{l c c c c c c c}
	\hline \hline
	Protocol & \, $d$ \, &  \,\, $e_b^\mathrm{max}$  \, \, & \, \, $e_b^\mathrm{exp}$ \, \, & \, \, $R(0)$ \,\,& \,\, $R^\mathrm{exp}$ \,\,&  \,\, Sifting \, \,& \, \,$R^\mathrm{exp} \times$ Sifting  \\
	\hline 
	Chau15 & 4 & 50~\% & 0.778~\% & 1& 0.8170  & 1/6 &0.1362  \\
	 & 8 &  50~\% & 3.11~\% & 1 & 0.8172 & 1/28 & 0.0292 \\
	BB84 & 2 & 11.00~\% & 0.628~\% & 1 & 0.8901 & 1/2 -- 1${}^{*}$ & 0.4451 -- 0.8901 \\
	 & 4 & 18.93~\% & 3.51~\% & 2 & 1.4500  & 1/2 -- 1${}^{*}$ & 0.7250 -- 1.4500\\
	 & 8 & 24.70~\% &10.9~\% & 3 & 1.3942  & 1/2 -- 1${}^{*}$ & 0.6971 -- 1.3942 \\
	MUB & 2 & 12.62~\% & 0.923~\% & 1 & 0.8727  & 1/3 -- 1${}^{*}$ & 0.2909 -- 0.8727 \\
	 & 4 & 23.17~\% & 3.87~\% & 2 & 1.5316  & 1/5 -- 1${}^{*}$ & 0.3063 -- 1.5316 \\
	Singapore & 2 & 38.93~\% & 1.23~\% & 0.4 & 0.374$^{**}$ & 1 & 0.374$^{**}$\\
	 \hline \hline
	\end{tabular}
\caption[]{Quantum bit error rates and key rates are presented for various quantum key distribution protocols. Four protocols, in various dimensions $d$, were investigated. The theoretical values of the error-free secret key rate, i.e. $R(0)$, and the maximum QBER, i.e. $e_b^\mathrm{max}$ for which $R=0$, are presented for the different protocols alongside the experimentally measured QBER, $e_b^\mathrm{exp}$, and secret key rates, $R^\mathrm{exp}$. Finally, the sifting rate, defined as the probability of obtaining sifted data, is also shown for each protocols. {${}^{*}$Depending on the size of the key, the choice of basis can be biased to get a larger sifting efficiency than 1/2 and 1/(d+1) for BB84 and MUB, respectively~\cite{tomamichel:12}. In the infinite key limit, the sifting efficiency can be made to approach 1.} ${}^{**}$Experimental rate for the Singapore protocol is deduced based on the point that a rate of 0.4 per 0.415 value of mutual information can be achieved. }
\label{table:1}
\end{table*}

The channel can be characterized as a positive trace-preserving map $\mathcal{E}$ such that ${\hat \rho}_{\text{out}}=\mathcal{E}({\hat\rho}_{\text{in}})$. This can then be described by the $d^2\times d^2$ process matrix, $\chi$, where $\mathcal{E}({\hat \rho}) = \sum_{ij}\chi_{ij} {\hat\sigma}_i {\hat\rho} {\hat\sigma}_j^\dagger$. In $d=2$, ${\hat\sigma}_i$ are identity and Pauli matrices. This approach can be extended to to higher dimensions using Gell-Mann matrices as they also offer an orthogonal basis, $\text{Tr}({\hat\sigma}_i{\hat\sigma}_j)=2\delta_{ij}$, spanning the vector space of complex matrices. {Moreover, the trace-preserving assumption may be eliminated while using a similar computational approach to reconstruct the process matrix~\cite{bongioanni:10}. By doing so, we may take into consideration mode-dependent losses coming from our measurement scheme or the quantum channel itself. Mode-dependent loss is insignificant in our laboratory-scale quantum channel, but may become an issue on longer distance links~\cite{sit:17}.}

 Given the experimental preparation and measurement results, we parametrize the process matrix and minimize a maximum likelihood function of
\begin{eqnarray*}
 && \ \ f (\vec{t} ) =  \\
&& \sum_{a,b}\frac{[ N_{ab}/N - \langle\psi_b| (\sum_{i,j}\chi_{ij} (\vec{t} ) {\hat\sigma}_i |\psi_a\rangle\langle\psi_a|{\hat\sigma}_j )|\psi_b\rangle ]^2}{2 \langle\psi_b| (\sum_{i,j}\chi_{ij} 
 (\vec{t} ) {\hat\sigma}_i |\psi_a\rangle\langle\psi_a|{\hat\sigma}_j )|\psi_b\rangle},
\end{eqnarray*}

to find the process matrix. Here $N_{ab}/N$ are normalized prepare-and-measure results, and $|\psi_a \rangle$ ($|\psi_b\rangle$) are prepared states (measurement projection settings). Using numerical minimization, we find the process matrix for both the six-state and Singapore protocols that are depicted in Fig.~\ref{fig:MUBSing}b. The quality of the channel may be described using the process fidelity, which is defined as $F=\mathrm{Tr} \left[ \chi_\mathrm{exp}\, \tilde{\chi} \right]$, where $\chi_\mathrm{exp}$ is the experimentally reconstructed process matrix and $\tilde{\chi}$ is the ideal process matrix, {i.e. $\tilde{\chi}_{i,j}=\delta_{i,0} \delta_{j,0}$, where $\delta_{i,j}$ is the Kronecker delta.} The process fidelity obtained from the process tomography using MUB and SIC-POVMs are given by $F_\mathrm{MUB}=98.7~\%$ and $F_\mathrm{SIC-POVM} = 95.8~\%$, respectively. Although SIC-POVMs offer a more efficient tomography of the channel, measurements of modes belonging to OAM MUB are of higher quality, leading to a larger process fidelity. This capability in tomographic protocols can potentially be used to identify attacks, and pre- or post-compensate for non-dynamical errors in the channel~\cite{bouchard:18c}.

\section{Conclusion}
Application of quantum physics in public key cryptography first emerged in the seminal work of Bennet and Brassard in 1984; leading to several other protocols that benefit from different properties of quantum states for secure quantum communications. Many experimental efforts have been dedicated to physical implementation of these protocols using mostly polarization and temporal degrees of freedom of photons. Despite significant progress in experimental realization of QKD protocols, each demonstration is practically limited to implement a single protocol in a specific dimension. Structured light, on the other hand, have been shown to offer flexibility in quantum state preparation and measurement in a theoretically unbounded Hilbert space. Here, we employed the versatility offered by the OAM states of photons to perform an experimental laboratory survey of four classes of QKD protocols in different dimensions. Table \ref{table:1} summarizes the main results of the several QKD schemes. This included  the Chau15 scheme (in $d=4$, and 8) based on differential phases, the BB84 protocol in dimensions 2, 4, and 8, and tomographic protocols based on $(d+1)$-MUB in $d=2$ (six-state), and 4, and SIC-POVMs in $d=2$. We observed experimental secure bit rates that ranges from 0.03 to 0.72 bit per sifted photon with schemes that have error tolerances from 11~\% up to 50~\%. {In particular, for the case of 2-dimensional tomographic protocols, for similar noise levels, the Singapore protocol can outperform the six-state protocol after sifting. For higher dimensions, the cross-talk among the different OAM modes limits the performance of several protocols. For example, the 8-dimensional BB84 fails to offer any advantage except for higher error tolerance compare to its 4-dimensional counterpart. However, there is a clear benefit in using the 4-dimensional BB84 rather than the 2-dimensional BB84 protocol, both in terms of key rate and noise tolerance given our experimental configuration. Moreover, in the case of the Chau15 scheme, the sifting rate scales unfavourably with dimensions. Thus, it is likely that under most conditions the 4-dimensional Chau15 scheme will be optimal.} Our experimental setup allows one to easily switch between protocols and dimensions to benefit from advantages of different protocols under varying channel conditions. This included using tomographic protocols for a more elaborate characterization of the errors in the quantum channel.

\noindent
\vspace{0.5 EM}

\noindent\textbf{Acknowledgments}
All authors would like to thank Gerd Leuchs, Markus Grassl and Imran Khan for helpful discussions. F.B. acknowledges the financial support of the Vanier graduate scholarship of the NSERC. R.F. acknowledges the financial support of the Banting postdoctoral fellowship of the NSERC. This work was supported by Canada Research Chairs; Canada Foundation for Innovation (CFI); Canada Excellence Research Chairs, Government of Canada (CERC); Canada First Research Excellence Fund (CFREF); Natural Sciences and Engineering Research Council of Canada (NSERC); Singapore Ministry of Education (partly through the Academic Research Fund Tier 3 MOE2012-T3-1-009) and the National Research Foundation of Singapore; and Spanish Ministerio de Economia y Competitividad (MINECO).


\bibliographystyle{plainnat}

\onecolumn\newpage
\appendix

\section{Details on estimating mutual information based on experimental results of SIC POVMs}
The performance of quantum key distribution protocols based on SIC POVMs can be assessed based on the mutual information between \emph{Alice} and \emph{Bob}. We use the experimental joint probability distribution that is depicted in Fig.~\ref{fig:BB84}(c) and (f) to find the experimetnal preparation and measurement vectors of \emph{Alice} and \emph{Bob}. For this, we parametrize $|\psi_m^{A,B}(x)\rangle$ vectors (as \emph{Alice}'s and \emph{Bob}'s experimental preparation and measurement states). Then, we numerically minimize a maximum likelihood function of $f(x)=\sum_{m,n=1}^{d^2}{\left| |\langle\psi_m^A(x)|\psi_n^B(x)\rangle|^2-|\langle\psi_m|\psi_n\rangle|^2\right|^2}$ to find the experimental SIC states of \emph{Alice} and \emph{Bob} with respect to the experimental joint probability distribution. The Singapore protocol relies on an anticorrelation between \emph{Alice} an \emph{Bob}. In the entanglement-based version of this protocol, this can be achieved by sharing a singlet entangled state. Assuming a singlet states of $|\Psi^{(-)}\rangle_{AB}=\frac{1}{\sqrt{d}}\Sigma_{m=1}^{m=d}(-1)^{d-m}|m\rangle|d-m\rangle$, we can extract the joint anti-correlated measure-and-prepare probailities of $p_{kl} = Tr(\rho_{AB}(1+t_k.\sigma_A)(1+t_l.\sigma_B))$, where $t_k$s are unit vectors denoting SIC states. In dimension d=2, this approach leads to 
\begin{equation}
P_{\text exp} =
\begin{bmatrix}
   0.000685 & 0.086229 & 0.067171 & 0.080468\\
   0.080968 & 0.000121 & 0.080109 & 0.072486\\
   0.091623 & 0.097436 & 0.001494 & 0.084059\\
   0.080614 & 0.093169 & 0.082246 & 0.001120\\
\end{bmatrix}, 
\end{equation}
compared to the ideal case of $p_{kl} =\frac{1-\delta_{kl}}{12}$. The mutual information between \emph{Alice} and \emph{Bob} is given by
\begin{equation}\label{I_AB}
I_{AB} = \Sigma_{k,l=1}^{d^2} p_{kl}\log_2\frac{p_{kl}}{p_{k\cdot}p_{\cdot l}},
\end{equation}
where $p_{k\cdot} = \Sigma_{l=1}^{d^2}p_{kl}$ and $p_{\cdot l} = \Sigma_{k=1}^{d^2}p_{kl}$. In $d=2$, the associated mutual information is $I_{AB}^{\text exp}= 0.408$ compared to the ideal mutual information of 0.415; see~\cite{englert:04}.

\end{document}